\documentclass{scrartcl}
\usepackage[utf8]{inputenc}
\usepackage[dvipsnames]{xcolor}
\usepackage{url}
\usepackage{svg}

\newcommand{\NEB}{\ensuremath{\mbox{NEB}}}
\newcommand{\IRV}{\ensuremath{\mbox{IRV}}}

\DeclareOldFontCommand{\tt}{\normalfont\ttfamily}{\mathtt}
\DeclareOldFontCommand{\bf}{\normalfont\bfseries}{\mathbf}
\DeclareOldFontCommand{\it}{\normalfont\itseries}{\textit}

\usepackage{authblk}

\title{You can do RLAs for IRV}
\subtitle{The Process Pilot of Risk-Limiting Audits for the San Francisco District Attorney 2019 Instant Runoff Vote}

\author{Michelle Blom\thanks{School of Computing and Information Systems, University of Melbourne.  \\ \texttt{michelle.blom@unimelb.edu.au}},
 Andrew Conway\thanks{Silicon Econometrics Pty. Ltd., \texttt{andrewelections@greatcactus.org}},
 Dan King\thanks{Viewpoint Technical, Chula Vista, California. \texttt{dan.king@sfgov.org, \texttt{dan.king@vptech.io}}},
 Laurent Sandrolini\thanks{San Francisco Elections Department, \texttt{laurent.sandrolini@sfgov.org}},
 Philip B. Stark\thanks{Department of Statistics, University of California, Berkeley.  \texttt{stark@stat.berkeley.edu}},
 Peter J. Stuckey\thanks{Department of Data Science \& AI, Monash University.  \texttt{Peter.Stuckey@monash.edu}}
and Vanessa Teague\thanks{Thinking Cybersecurity Pty. Ltd.  \texttt{vanessa@thinkingcybersecurity.com} \\Authors listed alphabetically.}
}

\date{\today}

\usepackage{graphicx}

\begin{document}

\maketitle

\begin{abstract}
    The City and County of San Francisco, CA, has used Instant Runoff Voting (IRV) for some elections since 2004.  This report describes the first ever process pilot of Risk Limiting Audits for IRV, for the San Francisco District Attorney's race in November, 2019.  We found that the vote-by-mail outcome could be efficiently audited to well under the 0.05 risk limit given a sample of only 200 ballots.  All the software we developed for the pilot is open source.
\end{abstract}

\section{Introduction}

Post-election audits test a reported election result by randomly sampling paper ballots.\footnote{%
 We use the terms ``ballot,'' ``ballot card,'' and ``card'' as synonyms, even though a ballot might comprise more than one physical card.
 Election audits generally sample cards rather than ballots:
 most voting systems cannot identify separate cards that comprise a single voter's multi-card ballot.
 }
A \emph{Risk Limiting Audit (RLA)} of a trustworthy paper trail of votes
either finds strong statistical
evidence that the reported outcome is correct, or reverts to a full manual tabulation to set the record straight.\footnote{%
  A careful, accurate full hand count finds the correct winner(s) if the paper trail is trustworthy---which is not automatic.
  }
(The outcome is the political result---i.e., who won---not the exact vote counts.)
The maximum chance that a RLA will fail to correct the reported outcome if the
reported outcome is wrong is the \emph{risk limit}.
RLAs are becoming the
\emph{de facto} standard for 
post-election audits that check the tabulation. 
They are required by statute in Colorado, Nevada, Rhode Island, and Virginia, and have been piloted in over a dozen US states and in Denmark.
California AB2125 authorizes RLAs.

\emph{Instant Runoff Voting (IRV)} allows voters to express their preference order (ranking) for some or all candidates.
IRV elections are counted by iteratively eliminating the least-popular candidate, as described in Figure~\ref{alg:IRV}.  
When a candidate is eliminated, each of their votes is passed to the next-preferred candidate on each ballot.  
The winner is the last remaining candidate when all the others have been eliminated. 
IRV is the normal form of voting in Australia, and is used or will be used in numerous US counties including San Francisco, Aspen, Oakland, and New York.

\begin{figure}[t]
	\centering
	\begin{tabbing}
		xx \= xx \= xx \=\kill
		Initially, all candidates remain standing (are not eliminated)\\
		\textbf{While} there is \textit{more than one} candidate standing \\
		\> \textbf{For} every candidate $c$ standing\\
		\> \> Tally (count) the ballots in which $c$ is the highest-ranked \\
		\>\> \> candidate of those standing\\
		\> Eliminate the candidate with the smallest tally\\
		The winner is the one candidate not eliminated
	\end{tabbing}
	\caption{The IRV counting procedure.}
	\label{alg:IRV}
\end{figure}

RLAs have been conducted for a variety of social choice functions (plurality, majority, super-majority, multi-winner plurality) but never for IRV.  
These can be audited by statistical tests of simple assertions about the ballots such as candidate A getting more votes than candidate B.
The complexity 
of IRV introduces challenges for RLAs because it may not be clear what assertions about the election need to be audited.  
In some IRV races, the only contest that really matters is the comparison between the last two uneliminated candidates; in others, a change in the early stages of the elimination sequence can cascade into a different election outcome.
The San Francisco pilot relied on theory derived only recently by Blom et al. \cite{blomEtal19} for analyzing the cast votes records (CVRs) to determine a set of simple, auditable, assertions which, taken together, imply that the reported election outcome is correct.

\subsection{Overview of the San Francisco DA pilot audit}
The San Francisco RLA pilot audited the vote by mail ballots for the 2019 San Francisco District Attorney's race.  Obviously, auditing only the votes cast by mail does not truly test the accuracy of the election outcome.  
In this case it happened that the outcome for the vote by mail ballots was different from the overall outcome---Susan Loftus won the vote by mail ballots quite comfortably, though Chesa Boudin won the election overall.  Hence the audit itself does not actually prove anything about the overall winner. Instead, it tested whether Susan Loftus would have won if the vote by mail ballots had been the only ballots cast.  Nevertheless it makes an interesting case study with which to explain how the general IRV RLA process works.

We call this a ``process pilot'' because it tested the feasibility of the process, not the election result itself.  
It was not a true RLA in part because it considered
only ballots cast by mail, since the voting system was able to match paper ballots to CVRs for those ballots but not for ballots cast in person.  Nevertheless, it gives us a good 
estimate of the amount of work that would be required to administer a meaningful RLA of an election with similar parameters.  
The audit required a sample of only 200 ballots even though the margin was small, and terminated with an estimated risk of only 0.003, well under the 0.05 risk limit.  
With three pairs of people entering the ballot data, the elapsed time for the audit was less than one hour, not including the time required to retrieve the paper ballots.  

This encourages optimism that RLAs for IRV is feasible, particularly when individual ballots can be compared with their CVRs.
It dispels the previously common but mistaken belief that IRV audits should take longer than audits of simpler voting systems.
They don't; they're just a little harder to understand.  
Any audit can require inspecting many ballots when the margin is close or the error rate is high, but there is no evidence that 
IRV audits are likely to require substantially more work than audits of other social choice functions.

Section~\ref{sec:pilotToReal} contains a discussion of how to extend the process pilot to a full election audit.

\subsection{The Software} \label{subsec:SoftwareSummary}
Two important new ideas were put into practice for the first time for this pilot.  
The first was the RAIRE 
IRV assertion generator, which turns a complete set of IRV CVRs into a set of simple assertions that can be tested by existing RLA methods.  The second was the SHANGRLA auditing framework, which presents a very general and flexible interface for RLAs and can incorporate RAIRE's assertions as well as assertions for other voting methods such as Borda, Condorcet, STAR-Voting, multi-winner plurality, and super-majority.
The audit also used a new ``risk-measuring'' function, the Kaplan Martingale (KMart).

The project produced five main pieces of software, all open source and easily available online:
\begin{description}
\item[A format converter and election counter] reads the CVRs and counts the votes to check that the outcome implied by the CVRs matches the reported election outcome.

\url{https://github.com/pbstark/SHANGRLA/blob/master/ConvertCVRToRAIRE.html}
\item[The RAIRE Assertion-generator] inputs the reformatted CVRs and calculates a set of assertions which, if true, imply that the reported election outcome is right.  RAIRE uses heuristics to choose assertions that can be audited efficiently.  See Section~\ref{subsec:RAIRE} for an explanation and Blom et al. \cite{blomEtal19} for more detail.

\url{https://github.com/michelleblom/audit-irv-cp/tree/raire-branch}

\item[The IRV assertion visualiser] displays a visual representation of all possible IRV election outcomes, allowing auditors to check directly that the assertions generated by RAIRE are sufficient to prove the reported election outcome.

\url{https://github.com/pbstark/SHANGRLA/blob/master/Code/RAIREExampleDataParsing.ipynb}

\item[The SHANGRLA RLA tool] is a general tool for conducting RLAs involving complex elections and a variety of possible statistical tests.  
It inputs the assertions from RAIRE and constructs assertions for other social choice functions (e.g., plurality, multi-winner plurality, or super-majority) and administers the audit.  
See Section~\ref{subsec:SHANGRLA} for an explanation and \cite{starkSHANGRLA} for more detail.

\url{https://github.com/pbstark/SHANGRLA}

\item[The Manual Ballot Entry Tool] inputs the list of randomly-selected ballot cards for audit, and allows the auditors to record what they see on the ballot.  
This information is then fed back into SHANGRLA to decide whether the audit
can stop or must examine more ballot cards.

\url{https://github.com/dan-king/RLA-MVR}
\end{description}

\section{How the software works}

Here we show how to adapt existing RLAs to IRV.  
The key insight is that we don't have to verify all the complicated steps of an IRV count---we find a few simple assertions that imply that the election outcome is right, then conduct an audit to test whether those assertions are true.
If the RLA doesn't find sufficiently strong evidence that those assertions
are true, the audit eventually expands to a full manual tabulation.

Before any votes have been tallied, we can imagine all possible elimination sequences $l_1, l_2, \ldots, l_{k}, w$, meaning that $l_1$ is eliminated first, followed by $l_2$, etc., in sequence, until $l_k$ and $w$ are the last two candidates standing, and $l_k$ has fewer votes (and is therefore eliminated).  
The last candidate in the list is the winner---the one who remains after everyone else has been eliminated.  Without knowing anything about the votes, we know that if there are $k+1$ candidates there must be $(k+1)! = (k+1) \times k \times (k-1) \times \ldots \times  3 \times 2$ different possible elimination orders.

\begin{figure}
    \centering
    \includegraphics[trim=15 0 0 400, clip, scale=0.8]{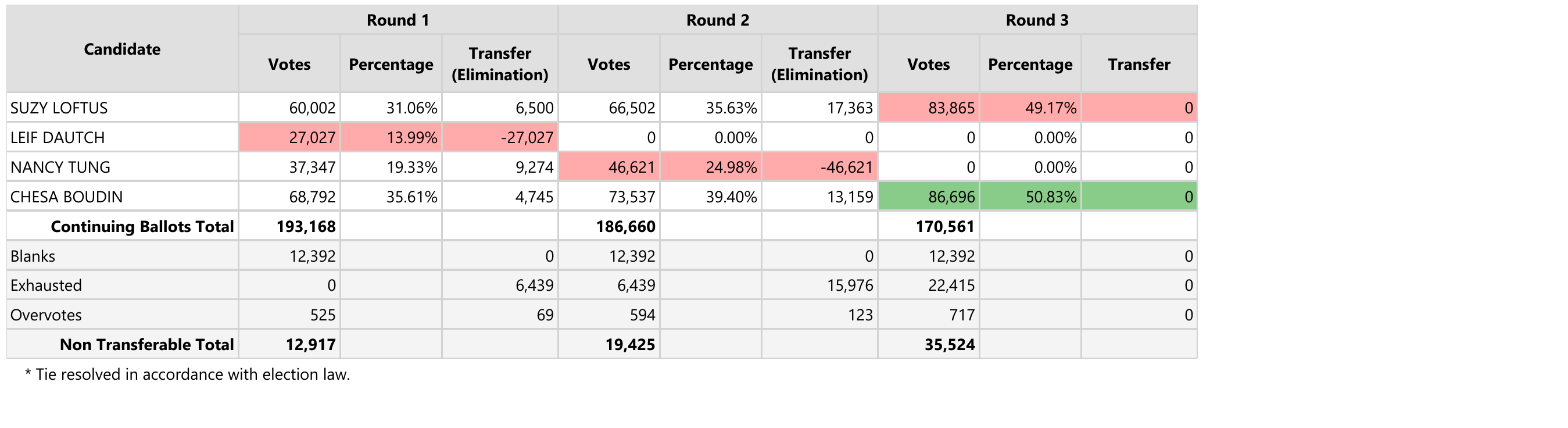}
    
    \caption{Official results, including elimination order, for the San Francisco DA race}
    \label{fig:officialResults}
\end{figure}

\begin{figure}
    \centering
    \includegraphics[trim=70 150 0 60, clip, scale=0.9]{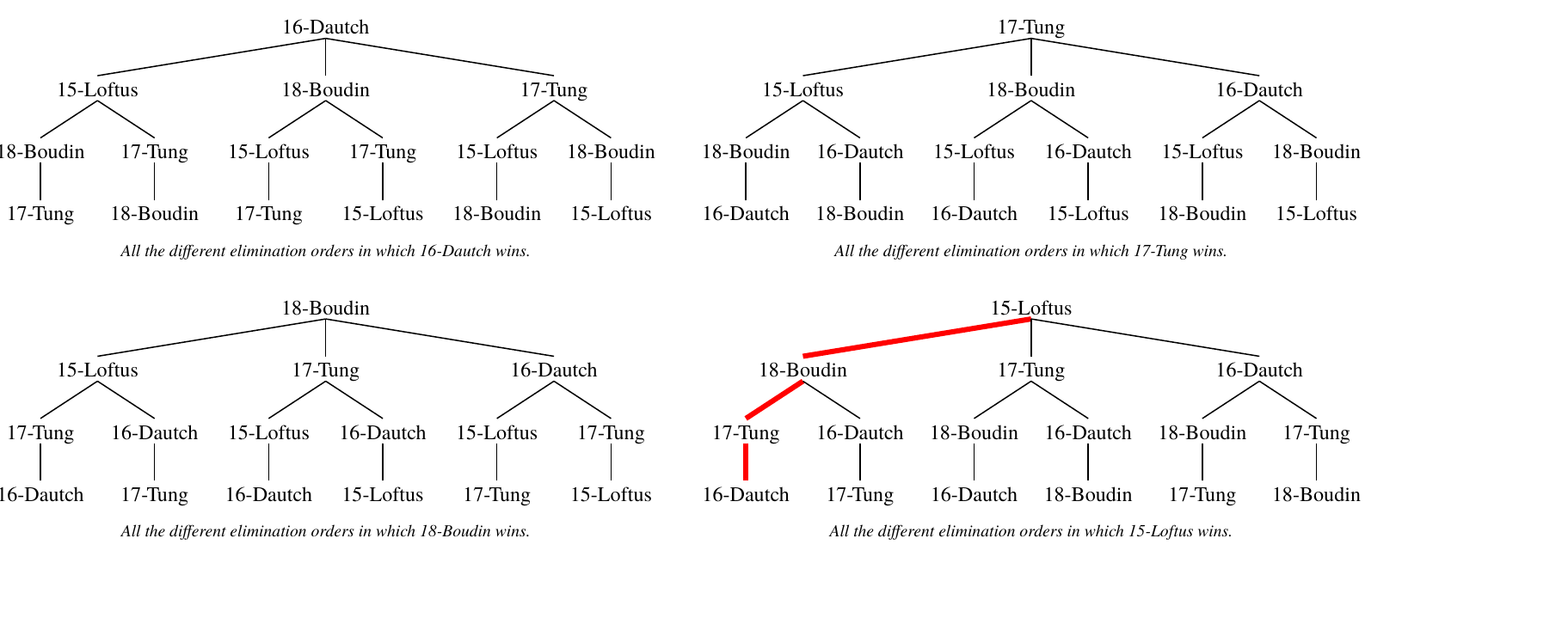}
    
    \caption{Complete Elimination Trees for the San Francisco DA race}
    \label{fig:SFDACompleteTrees}
\end{figure}

These $(k+1)!$ elimination orders can be arranged into $k+1$ trees, one for each winning candidate.  
The root of each tree is the winner, while each path from a leaf to the root represents a possible elimination order, with the first-eliminated candidate at the leaf, the next eliminated candidate as its parent node, and so on.
In the San Francisco DA race, the apparent elimination order (for VBM ballots) was Dautch, Tung, Boudin, Loftus---this is shown in Figure~\ref{fig:officialResults}.
An example of the complete list of elimination trees is shown in Figure~\ref{fig:SFDACompleteTrees}, with the reported election outcome marked in red.  
Other paths in the same tree also represent wins for Loftus, by different elimination sequences.

First observe that to test whether Loftus truly won there is no need to check the exact elimination sequence.  
If the reported winner truly won, but by a different elimination sequence, the reported election outcome is still correct.  
Therefore there is no need to audit anything about the tree of possible ways in which Loftus won.  
Instead, we concentrate on checking that no elimination sequence with a different winner is possible---visually, this corresponds to pruning every other tree so that every path from a leaf to the root is broken somewhere by an assertion that can be excluded by auditing.

The next section explains how RAIRE constructs assertions
that perform this pruning.

\subsection{Overview of RAIRE} \label{subsec:RAIRE}

Sometimes the only way to audit an IRV election is to to check that, at every step of the process, the right candidate was eliminated.  
In general, however, that strategy is inefficient because it may take a lot of work to verify comparisons that don't matter---for instance if a change to the early elimination order makes no difference to the final result.  
The next sections describe quicker ways in which it is sometimes possible to be confident that the reported winner truly won, without checking the entire elimination sequence.

If every possible elimination sequence that produces a different winner (other than the reported winner $w$) can be contradicted by a true assertion, then every other possible winner has been 
excluded and $w$ really won.
RAIRE produces a set $\mathcal{F}$ of assertions such that \emph{if} all the assertions in $\mathcal{F}$ are true, \emph{then} $w$ truly won.
We conduct the overall RLA by checking each assertion in $\mathcal{F}$
as if it were the reported outcome of a 2-candidate plurality contest.  
The same sample can be used to check all the assertions.

RAIRE generates the assertions that can be used to prune the tree, but it is not necessary to trust RAIRE to do this correctly.  The tree visualisation software allows any observer to check for themselves that every tree in which some candidate other than the reported winner wins has been completely pruned.  
Figures~\ref{fig:SFDAInitialTree} and~\ref{fig:SFDA11Tree} show examples of tree visualisations for the San Francisco DA race---you can check for yourself that there is no remaining unpruned path from a leaf all the way to the root.

\subsubsection{``IRV-elimination'' assertions} \label{subsubsec:elimSeq}

Suppose that one branch we wish to prune is an elimination sequence $l_1,\ldots,l_k$ with candidate $w'$ the (alternative) winner.  
If $w'$ is not the true winner, there must be at least one step along this sequence of eliminations that we can rule out.
Consider the $r$-th step, in which $l_r$ is eliminated.  
This elimination step is like a multi-winner
plurality (first-past-the-post) election with one loser ($l_r$) and $k-r+1$ winners $l_{r+1},\ldots,l_k,w$.  
We disregard all the candidates that have already been eliminated ($l_1,\ldots,l_{r-1}$) and attribute each ballot to whichever candidate in the set $l_r,l_{r+1},\ldots,l_k,w$ it ranks highest.  
RAIRE can prune this branch by checking the assertion that that $l_r$ must beat one of $l_{r+1},\ldots,l_k,w$ at this step.

\begin{quotation}
$\IRV{}(l_r, c, \{ l_{r+1},\ldots,l_k,w \})$ is the assertion that $l_r$ beats $c \in \{ l_{r+1},\ldots,l_k,w \}$ when each ballot card is counted as a vote for the candidate in $l_r,l_{r+1},\ldots,l_k,w$ ranked highest on that card.
\end{quotation}

\begin{figure}
    \centering
    \includegraphics[trim=0 75 0 70, clip, scale=0.8]{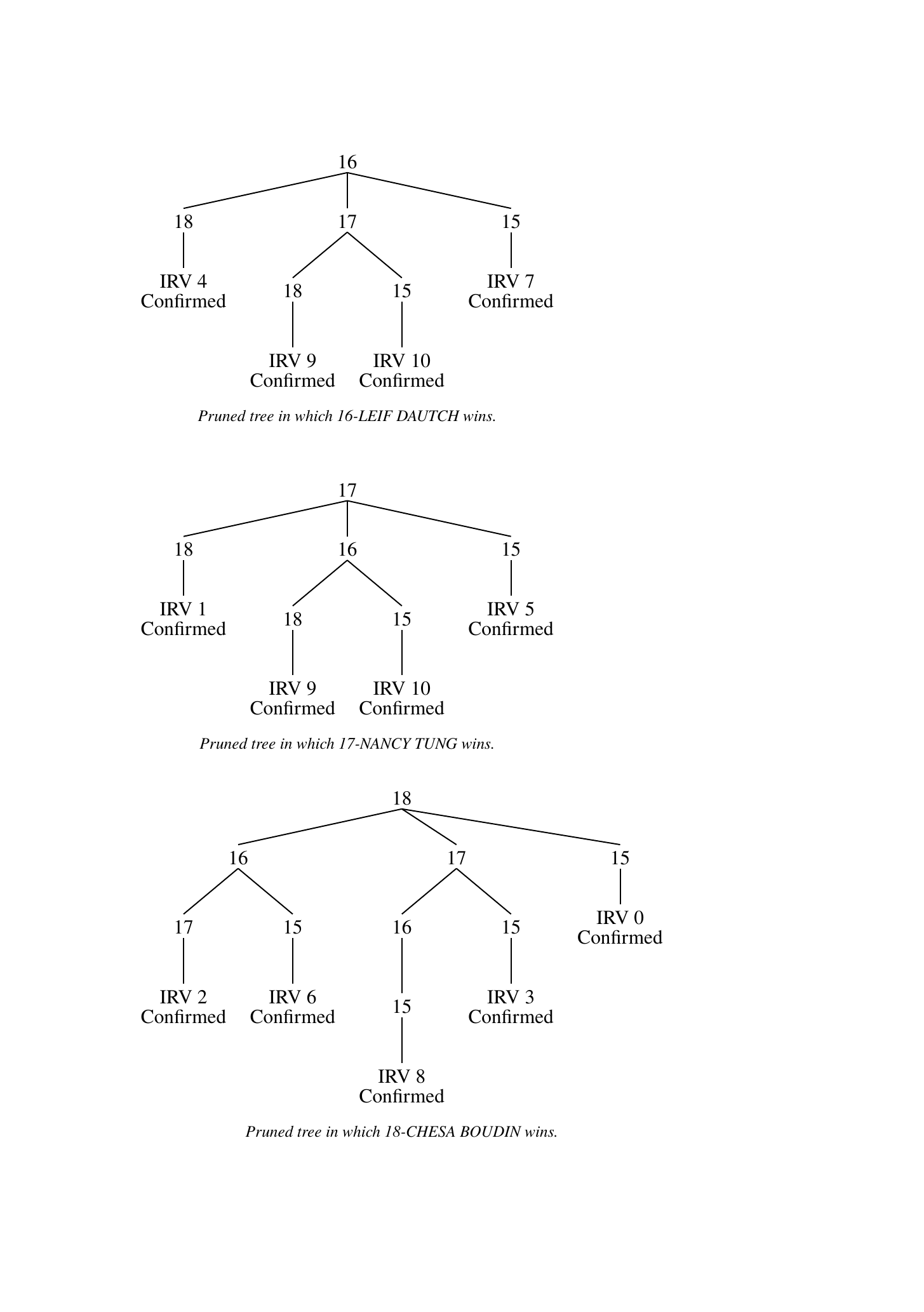}
    
    \caption{Pruned Elimination Trees used for the San Francisco DA RLA. 
    IRV $n$ means that section
    of the tree is impossible if the assertion IRV $n$ (listed in figure \ref{fig:InitialAssertionsExpl}) is true. Confirmed means that the
    audit has confirmed that assertion.}
    \label{fig:SFDAInitialTree}
\end{figure}

The visualisation of alternative trees for the San Francisco DA race is shown in Figure~\ref{fig:SFDAInitialTree}.  
Note that every branch of every tree (other than the tree in which Loftus wins) is pruned by an IRV assertion.
The explanation of each assertion is shown in Figure~\ref{fig:InitialAssertionsExpl}.


\begin{figure}
\small
\noindent \textbf{IRV assertions:} \\ 
Confirmed:   IRV  0: Candidate 15 cannot be eliminated next when \{16, 17\} are eliminated. \\
Confirmed:   IRV  1: Candidate 18 cannot be eliminated next when \{16, 15\} are eliminated. \\
Confirmed:   IRV  2: Candidate 17 cannot be eliminated next when \{15\} are eliminated. \\
Confirmed:   IRV  3: Candidate 15 cannot be eliminated next when \{16\} are eliminated. \\
Confirmed:   IRV  4: Candidate 18 cannot be eliminated next when \{15, 17\} are eliminated. \\
Confirmed:   IRV  5: Candidate 15 cannot be eliminated next when \{16, 18\} are eliminated. \\
Confirmed:   IRV  6: Candidate 15 cannot be eliminated next when \{17\} are eliminated. \\
Confirmed:   IRV  7: Candidate 15 cannot be eliminated next when \{18, 17\} are eliminated. \\
Confirmed:   IRV  8: Candidate 15 cannot be eliminated next when \{\} are eliminated. \\
Confirmed:   IRV  9: Candidate 18 cannot be eliminated next when \{15\} are eliminated. \\
Confirmed:   IRV 10: Candidate 15 cannot be eliminated next when \{18\} are eliminated. \\

\caption{Explanation of assertions for the Elimination Trees of Figure~\ref{fig:SFDAInitialTree}.}
\label{fig:InitialAssertionsExpl}
\end{figure}

\subsubsection{``Not-eliminated-before'' assertions}
``Not-eliminated-before'' auditing is a surprisingly powerful technique for proving that a certain candidate cannot win.  
It compares the highest possible tally of a reported loser to the lowest possible tally of the
reported winner.  
The lowest tally that $w$ can possibly have at any elimination stage is its total number of first preferences---IRV adds but never subtracts votes from not-yet-eliminated candidates as the algorithm progresses.  
The highest tally loser $l$ can possibly have (assuming $w$ is not eliminated) is the total number of mentions of $l$ at any preference, when there is no higher preference for $w$ on the same ballot card.  
If $w$'s first preferences are greater than $l$'s total mentions (excluding the ones listed below $w$), then $l$ can never achieve a tally as large as $w$'s.  Therefore $w$ cannot be eliminated before $l$ in any elimination sequence.  

We call this hypothesis \emph{Not-Eliminated-Before}, $\NEB{}(l,w)$.  (It is called Winner-only auditing in~\cite{blom2019risk}.)

\begin{quotation}
	
$\NEB{}(l,w)$ is the assertion that the number of cards
that have $w$ as the first preferences is greater than the total number of cards that mention $l$ and do not prefer $w$
to $l$. 
\end{quotation}

If this assertion is true, $w$ cannot be eliminated before  $l$, so $l$ cannot win.  This assertion is most often useful when $w$ is the reported winner of the election, but can sometimes be applied for other candidates too.
Sometimes the assertion $\NEB{}(l_i,w)$ is true for every reported loser $l_i$, which is enough to prove that $w$ won.\footnote{
This argument can be extended to consider minimum and maximum tallies given that a certain set of candidates has already been eliminated---see~\cite{blom2019risk} for details.}

\begin{figure}
    \centering
    \includegraphics[scale=0.8]{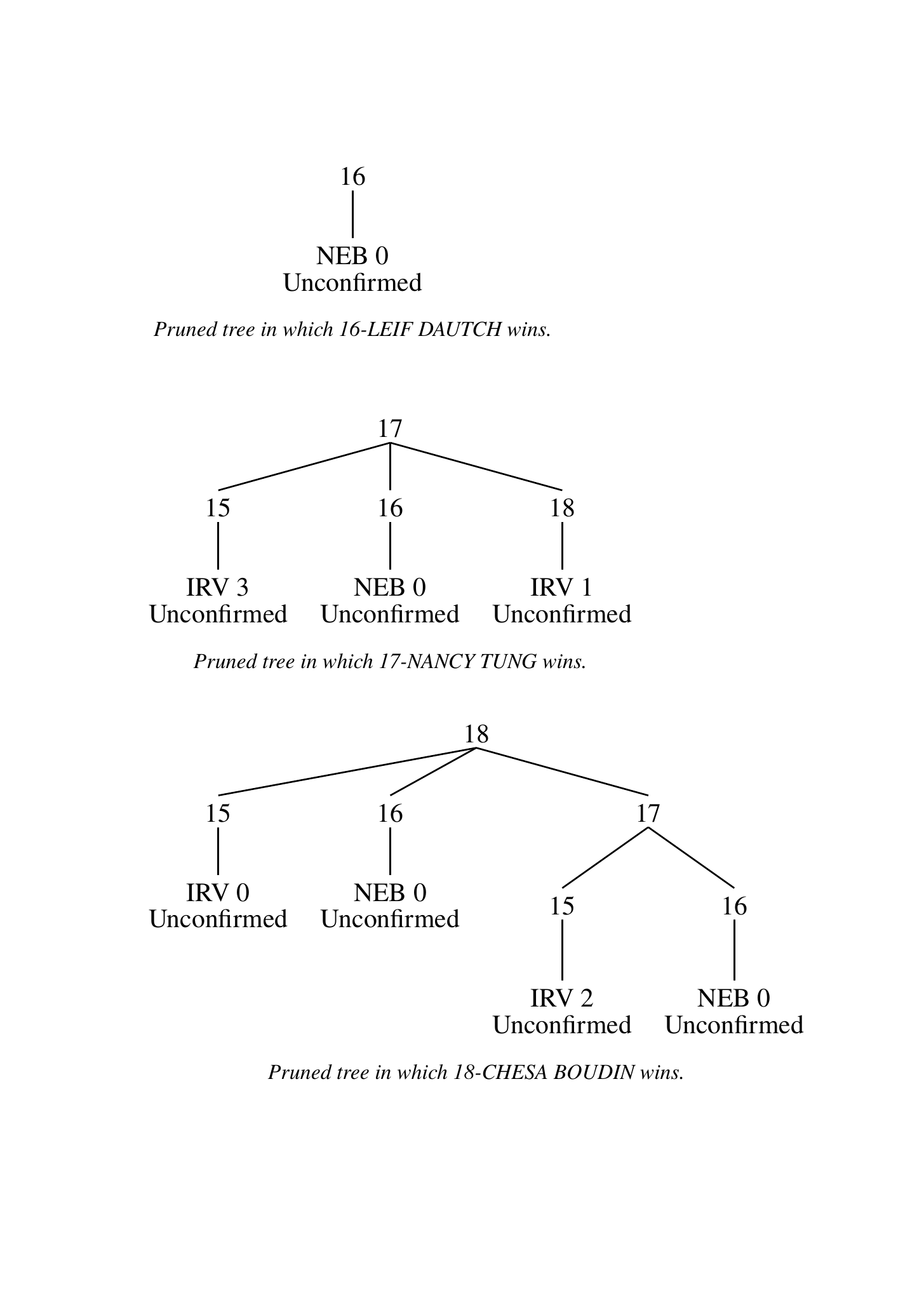}
    
    \caption{A valid alternative set of assertions for testing the outcome of the San Francisco DA race}
    \label{fig:SFDA11Tree}
\end{figure}

An example for an alternative method of auditing the San Francisco DA race incorporating an \NEB{} assertion is shown in Figure~\ref{fig:SFDA11Tree}, based on the 11th round of preliminary results.  For those preliminary results, candidate 16 (Dautch) could be excluded immediately by an  \NEB{} assertion (i.e. at least one other candidate could not be eliminated before her).  The assertions are all unconfirmed because this collection of assertions was never tested---the set shown in Figure~\ref{fig:SFDAInitialTree} was.  However, \emph{if} these assertions had been checked, they would have provided an alternative valid way of confirming the election outcome. The assertions are explained in Figure~\ref{fig:Prelim11AssertionsExpl}. 

\begin{figure}
\small
\noindent \textbf{Not-Eliminated-Before assertions:} \\ 
\NEB{}   0: Candidate 15 cannot be eliminated before 16. \\
\noindent \textbf{IRV assertions:} \\ 
IRV  0: Candidate 15 cannot be eliminated next when \{16, 17\} are eliminated. \\
IRV  1: Candidate 18 cannot be eliminated next when \{16, 15\} are eliminated. \\
IRV  2: Candidate 15 cannot be eliminated next when \{16\} are eliminated. \\
IRV  3: Candidate 15 cannot be eliminated next when \{16, 18\} are eliminated.
\caption{Explanation of assertions for the Elimination Trees of Figure~\ref{fig:SFDA11Tree}.}
\label{fig:Prelim11AssertionsExpl}
\end{figure}

\subsubsection{Summary of What RAIRE does}
RAIRE takes the reported set of votes, computes the apparent winner $w$, and finds a collection $\mathcal{F}$ of assertions that imply $w$ won.  As described above, each assertion in $\mathcal{F}$ is either an IRV-elimination or NEB.  These assertions should then be audited with an RLA.  RAIRE uses heuristics to try to find the $\mathcal{F}$ most likely to terminate in a successful audit in the shortest time.  This assumes, of course, that the reported outcome is correct---if it is not, then at least one of the assertions in $\mathcal{F}$ must be false, and this should be detected by the RLA with probability at least $1-\alpha$.  
If the audit of any assertion $f \in \mathcal{F}$ does not support $f$, a full manual recount should be performed.

\subsection{Overview of SHANGRLA} \label{subsec:SHANGRLA}
SHANGRLA is a very general method of auditing a variety of election types, by expressing an apparent election outcome as a series of assertions.
Each assertion is of the form ``the mean of a list of non-negative numbers is greater than 1/2.''  For example, consider an election with only two candidates, A and B, in which A is the reported winner. We  test the assertion that ``of those ballots that contain one candidate selection, more than half chose A.'' This can be audited in SHANGRLA by counting a vote for A as 1, a vote for B as 0 and a blank ballot (or a ballot that selects both) as $1/2$.  Now A is the true winner of the election if and only if the mean of those numbers is greater than $1/2$.

Each assertion is tested using a sequential test of the null hypothesis that its complement holds, i.e. the hypothesis that the mean is in fact less than or equal to 1/2.  If all the null hypotheses are rejected, the election outcome is confirmed. If not, we proceed to a full manual recount. 
SHANGRLA incorporates several different statistical risk-measurement algorithms and extends naturally to plurality and super-majority contests with various election types including Range and Approval voting and Borda count.

SHANGRLA is specifically designed to support
auditing Instant Runoff Voting (IRV) using the RAIRE assertion-generator. 
RAIRE produces a set of assertions sufficient to prove that the reported winner truly won, then SHANGRLA interprets these as assertions of the form ``the mean of a list of non-negative numbers is greater than 1/2'' and tests those assertions.

SHANGRLA also implements the ``manifest phantoms to evil zombies'' approach of \cite{banuelosStark12} which allows the audit
to sample only cards with CVRs that contain particular contests, while ensuring that the risk limit is met even if the CVRs are wrong.
An upper bound on the number of ballot cards that
contain each contest under audit is required.
The ability to target the sample makes it possible to audit contests that are not on every ballot card---such as partisan primaries and contests
that are not jurisdiction-wide---much more efficiently.
This is especially helpful for small contests with small margins, where it avoids ``diluting'' the sample be ensuring every selected card is informative and avoids ``diluting'' the contest margin by limiting the population of ballots to those that (putatively) contain the contest.
See \cite{starkSHANGRLA} for additional detail.

SHANGRLA code is available at \url{https://github.com/pbstark/SHANGRLA}, with a detailed explanation by Stark~\cite{starkSHANGRLA}.

\subsubsection{Expressing IRV assertions in SHANGRLA}
Consider testing the assertion $IRV(l_r, c, \{ l_{r+1},\ldots,l_k,w \})$.
In this assertion, $l_r$ is treated as the winner and $c$ as the loser, so a vote with $l_r$ as the highest-ranked candidate in $\{l_r,l_{r+1},\ldots,l_k,w\}$ is counted as 1.  A vote with $c$ as the highest-ranked candidate in $\{l_r,l_{r+1},\ldots,l_k,w\}$ is counted as zero.  Anything else is counted as 1/2.  Thus $l_r$ beats $c$ at this point in the elimination sequence if and only if the mean of those numbers is greater than 1/2.

\subsubsection{Expressing NEB assertions in SHANGRLA}
Consider testing the assertion $\NEB{} (l,w)$.
To express this using SHANGRLA, count a first preference for $w$ as a `vote' for $w$, i.e., a value of 1.  
Count any mention of $l$ with no higher preference for  $w$ as a `vote' for $l$, i.e., 0.  
Anything else is worth 1/2.

\section{Completing the steps for a full audit} \label{sec:pilotToReal}

There are several generic steps necessary for a true audit that were omitted from the process pilot, such as a \emph{compliance audit} to ensure that the paper trail was trustworthy,
a public dice-rolling ceremony to generate the seed,
 and public retrieval of the paper ballots from storage.  Since these are universal necessities for any RLA, we do not detail them here---see, e.g.,
 \cite{lindemanStark12}, \cite{starkWagner12}, \cite{morrell2019knowing} and \cite{morrell2019knowing2} for instructions.

The main challenge in extending this pilot to a full, meaningful audit in San Franscisco is incorporating votes that were cast in precincts.  
An audit that considers only VBM ballots proves nothing about the overall election outcome---this was particularly obvious this year because the reported winner on VBM ballots was different from the reported overall winner of the DA race.

In San Francisco at present, ballots that are cast in the precinct are not amenable to a ballot-comparison audit, because the way they are stored electronically and physically does not allow an auditor to retrieve the paper ballot corresponding to a particular CVR.  So there are three options.

\begin{enumerate}
    \item \label{op:comp} Update the procedure for ballots cast in the precinct so that it is possible, without violating vote privacy, to link a particular CVR with its paper ballot. 
    \item It might also be possible to do batch-level comparison audits in a roundabout way: if the CVRs for physical batches are available (even if they can't be matched to specific ballots within the batch), one could use them to compute `tallies' for the assertions, then check the  tallies by hand if the batch is selected for audit.  \label{op:batches}
    \item \label{op:suite} Finally, it might be possible to combine RAIRE with the SUITE audit method \cite{ottoboni2018risk}, which allows auditing of ballots from two or more different strata, in this case ballot-comparison and ballot-polling.
\end{enumerate}

The first option will result in examining the fewest ballots when the reported outcome is correct, though it requires some manual setup work.  

The second option requires less setup but probably more auditing.  A nice feature is that it wouldn't require stratification: batches can be drawn with probability proportional to an error bound as described in Section~3 of~\cite{lindeman2018next}. 

Option~\ref{op:suite} requires more careful thought. 
RAIRE uses heuristics to generate a set of assertions that are likely to require the least auditing work, assuming there are no errors.  
These heuristics rely on an estimate of the expected sample size, which depends on the audit method being employed. 
SUITE does a complementary kind of optimization, choosing the most efficient ratio of sample probabilities in the different strata in order to minimize the expected audit cost.  
So SUITE can optimize for a given set of RAIRE assertions, and RAIRE can optimize given a particular choice of SUITE sampling ratios, but it is not obvious how to do the joint optimization to minimize overall expected sample size.  
Fortunately, this optimization affects efficiency but not soundness, and a suboptimal solution might still be quite efficient in practice.  
For example, we could instruct RAIRE to generate assertions as if it was doing a ballot polling audit, then use those assertions for both the ballot-polling and ballot-comparison strata, in the ratio determined by SUITE.
However, RAIRE might in these cases over-estimate how much auditing is needed or even fail to produce any assertions because it seems much too hard.

\section{Conclusion}
You can do RLAs for IRV, using the open source software described in this report.  

Our pilot took fewer than six person-hours of work, excluding the time to retrieve the paper ballots.  The vote-by-mail outcome could be audited with a sample of only 200 ballots even though the margin was small, and terminated with an estimated risk of only 0.003, well under the 0.05 risk limit.

IRV audits can be as efficient as audits for simpler social choice functions, though of course a larger sample will be required if the margin is small, the error rate is large, or there is no way to match CVRs with their corresponding paper ballot.

All of the software developed for the San Francisco DA pilot audit is openly available online.

\section{Acknowledgements}
Many thanks to  the San Francisco Department of Elections for their support of this project.  
Thanks also to Yuvi Panda for hosting Jupyterhub and Damjan Vukcevic for valuable discussions.

\bibliographystyle{plain}


\appendix

\section{FAQ} \label{sec:FAQ}
\begin{enumerate}
\item What is the risk limit of the IRV audit?  
	
	Answer: It inherits the risk limit from the RLAs conducted on each assertion in $\mathcal{F}$. If every assertion in $\mathcal{F}$ is audited with risk limit $\alpha$, then the overall RAIRE audit detects a wrong election outcome with probability at least $1-\alpha$.
	
	\item How much auditing work will we need to do?
	
	Answer: It depends on factors such as the election margin and the number of discrepancies between the real and reported ballots.  Even for ordinary first-past-the-post elections, RLAs can be very fast when the margin is large and there are no errors, or relatively time-consuming when the margins are close or there are significant differences between real and reported ballots.  RAIRE also follows this pattern.
		It also depends on whether ballot-polling or ballot-level comparison audits are chosen.  
	
	\item Is it possible to estimate in advance how much auditing will be needed?
	
	Answer: Yes, but only on the assumption of a certain rate of error, which can't be predicted without inspecting the ballots.  SHANGRLA provides estimated 
	sample sizes given an estimated error rate.
	
	\item Might RAIRE fall back to a full manual recount even when the reported outcome is correct?  Is this more likely than for first-past-the-post audits?
	
	Answer: Yes, any RLA might fail to certify the result, and fall back to a full manual recount, even when the reported result is correct.  RAIRE is more likely to do this than an otherwise-equivalent RLA on a first-past-the-post election of the same margin, because it conducts several simultaneous audits, any one of which might behave in this way.

		\item Can we inspect the software?
	
	Answer: Yes, all the code is available at the links given in Section~\ref{subsec:SoftwareSummary}.
	
	\item Do we need to trust the RAIRE software?

No, you don't need to trust the software in order to be convinced by the audit---you can inspect the assertions $\mathcal{F}$ using the visualiser and check that they imply that the reported winner truly won.

	However, you do need a version of the RLA computations that you trust.  There are many options---you can trust SHANGRLA or choose to reimplement your own.

	\item Do we need to know the margin?  Aren't margins hard to compute for IRV?
	
	Answer: The true margin in an IRV contest isn't obvious, though it
        can usually be computed in reasonable time~\cite{blom2016efficient}.
        It is often, but not always, half the difference between the last
        two candidates standing in the last round.  RAIRE does not
        explicitly use the margin to construct the auditing assertions, but
        a lower bound on the margin is implied.  Each assertion $f \in \mathcal{F}$ can be thought of as having its own margin, which is the number of votes that would need to be altered in order to make that assertion false.  The overall IRV election margin cannot be smaller than the smallest margin of any assertion in $\mathcal{F}$.
\end{enumerate}
\end{document}